\documentclass[12pt]{article}
\usepackage{amssymb,amsmath,amsfonts,amsthm,eurosym,graphicx,color,setspace,sectsty,comment,footmisc,pdflscape,subfigure,array}
\usepackage[comma,authoryear]{natbib}
\usepackage[colorlinks=true,urlcolor=blue,citecolor=blue,linkcolor=blue,bookmarks=true]{hyperref}
\usepackage[font=small,skip=0pt]{caption}
\usepackage{cmap} 
\usepackage{mathtext}
\usepackage{rotating}
\usepackage{array,makecell,booktabs,multirow}
\newcolumntype{C}[1]{>{\centering\arraybackslash}p{#1}} 
\usepackage{booktabs,multirow,makecell,array}
\usepackage{bbm}
\usepackage{adforn} 
\usepackage{alphalph}
\usepackage[titletoc]{appendix}
\usepackage{commath}
\usepackage{datetime}
\usepackage{placeins}
\usepackage{footnote}
\usepackage{pdfpages}
\usepackage{fancyhdr}
\usepackage{mathrsfs}
\usepackage{enumerate}
\usepackage{adjustbox}
\usepackage{booktabs}
\usepackage{epstopdf}
\usepackage{longtable}
\usepackage{multirow}
\usepackage{ragged2e}
\usepackage{tabularx}
\usepackage[para, flushleft]{threeparttablex}
\usepackage{float}
\usepackage{wrapfig} 
\usepackage[T2A]{fontenc} 
\usepackage[utf8]{inputenc} 
\usepackage[english]{babel}
\usepackage{pgf,tikz} 
\usepackage{pgfplots}
\usepackage{pgfplotstable}
\usetikzlibrary{shapes}
\usetikzlibrary{arrows, calc}
\usepackage{titlesec}
\usepackage{authblk}
\usepackage{indentfirst}
\usepackage{csquotes}
\usepackage[left=2cm,right=2cm, top=2cm,bottom=2cm,bindingoffset=0cm]{geometry}
\usepackage{url}
\usepackage{subcaption}
\usepackage{graphicx}
\usepackage{caption}
\usepackage{rotating}
\usepackage{xcolor}
\usepackage{array,booktabs,xurl}
\newcolumntype{Y}{>{\raggedright\arraybackslash}p{4.9cm}} 

\begin{document}

\title{Strategizing with AI: \\ Insights from a Beauty Contest Experiment}

\date{}

\author[1]{Iuliia Alekseenko}
\author[1,3,*]{Dmitry Dagaev}
\author[2]{Sofia Paklina}
\author[2,4]{Petr Parshakov}

\affil[1]{HSE University, Laboratory of Sports Studies}
\affil[2]{HSE University, International Laboratory of Intangible-Driven Economy}
\affil[3]{New Economic School}
\affil[4]{SKOLKOVO School of Management}
\affil[*]{Corresponding author. Email: ddagaev@nes.ru. Address: 121205, Russia, Moscow, Innovation Centre Skolkovo, Nobel street, 3.}

\maketitle

\begin{abstract}
A $p$-beauty contest is a wide class of games of guessing the most popular strategy among other players. In particular, guessing a fraction of a mean of numbers chosen by all players is a classic behavioral experiment designed to test iterative reasoning patterns among various groups of people. The previous literature reveals that the level of sophistication of the opponents is an important factor affecting the outcome of the game. Smarter decision makers choose strategies that are closer to theoretical Nash equilibrium and demonstrate faster convergence to equilibrium in iterated contests with information revelation. We replicate a series of classic experiments by running virtual experiments with large language models (LLMs) who play against various groups of virtual players. Our results show that LLMs recognize strategic context of the game and demonstrate expected adaptability to the changing set of parameters. LLMs systematically behave in a more sophisticated way compared to the participants of the original experiments. All LLMs still fail to identify dominant strategies in a two-player game. Our results contribute to the discussion on the accuracy of modeling human economic agents by artificial intelligence.      
\end{abstract}


~~~~\textbf{Keywords}: $p$-beauty contest, guess the number, dominant strategy, AI, LLM

~~~~\textbf{JEL codes}: C99, D90, C72

\doublespacing
\newpage
\section{Introduction}

The recent appearance of large language models (LLMs) has resulted in numerous attempts to substitute humans with generative agents in various settings (see, for example, \cite{Park2023}). The motivation is simple: in many economic activities, especially those involving codifiable, high-volume, or knowledge-intensive tasks, LLMs are increasingly cost-competitive and productivity-enhancing compared to human labor \citep{korinek, merali, kanazawa2025ai}, and such automation can reduce employment \citep{Acemoglu2018}. Still, at this point, it is not clear to what extent LLM can simulate human behavior. The revelation of differences between the decisions of the participants of economic experiments and LLMs' decisions in similar circumstances becomes an important challenge. \cite{Horton2023} replicates several classic experiments with LLM players and advocates the use of LLMs as models for ordinary economic agents. \cite{akata2023} reveal that LLMs underperform in the iterated games that require coordination such as the iterated Battle of Sexes. \cite{brookins2023} show that LLMs prefer fair decisions in a Dictator game, even more than human players do. \cite{goli2024} study the intertemporal preferences of LLMs and figure out that artificial players are less patient than humans.
In general, different LLMs have different peculiarities: some are very sensitive to game structure, others have issues with context framing \citep{lore2024}.

Our paper contributes to this strand of literature by studying the behavior of LLMs in the classic \textit{Guess the number} game which belongs to a wider class of $p$-beauty contest games. These $p$-beauty contest games are of particular importance because they arise in various industries where the profit of a firm or the payoff of an individual player depend on the median preferences of all economic agents. The behavior of short-term traders and the resulting asset pricing is regarded as a variant of a beauty contest game \citep{allen2006, cespa2015}. On the prediction markets, the desire to guess the most popular opinion is often considered as a behavioral driving force in addition to guessing the fundamental probabilities of an outcome \citep{marinovic2011}. The beauty contest auction is an important alternative to price-only mechanisms of allocating goods to the bidders \citep{yoganarasimhan2016}. Whereas algorithmic trading and betting strategies written by humans have long been used on the markets, the appearance of strategies generated by LLMs is a rather recent phenomenon. Investigation of the LLMs' performance in a $p$-beauty contest game (both absolute and relative to humans) would help to predict their performance in the above-mentioned markets. 

The \textit{Guess the Number} game tests the ability of a player to make a sequence of conclusions, and the outcome depends on their level of sophistication. The literature provides several important observations about the performance of heterogeneous players in various strategic settings. First, and most obvious, more advanced players reach better outcomes. The latter observation is especially notable in antagonistic pure games of skill. \cite{charness} revealed that the objective quality of moves selected by chess players increases with the player's skill. \cite{Levitt} considered a sequential game \textit{Race to 100} which is a pure game of skill. They conclude that the ability to perform backward induction leads to better results in the \textit{Race to 100} game. Second (and this is less obvious), the players pay attention to the opponents' quality. In the \cite{eichberger} experimental study, most of the participants (72\%) who play the simultaneous games of strategic complements or strategic substitutes feel that they can predict the actions of a game theorist better than the actions of a granny. The same large majority (72\%) prefer to play against a game theorist rather than against a granny. In the \textit{Prisoner's Dilemma} game, the high-ability players who learn that their partner is a low-ability one, decrease the level of cooperation \citep{lambrecht}.
In antagonistic games, more advanced human players demonstrate greater adaptability to competing environment. For example, in the centipede game, stronger chess players exploit the non-equilibrium play of weaker players \citep{palacios}. However, \cite{Levitt} show that the ability to perform backward induction is not related to ability to stop early in the centipede game and assert that, instead of a centipede game, the pure games of skill should be used to test the role of the level of sophistication. We conclude that in many strategic environments, the skill of the players affects the outcome by allowing them to choose better strategies and/or to show greater adaptability to the opponents. To what extent do LLMs behave like a human? To answer this question, we replicate a series of well-known experiments with human participants playing the \textit{Guess the Number} game by asking LLMs to play against the same groups of competitors. 

The rules of the \textit{Guess the number} game are as follows. A group of $n$ players simultaneously and independently choose a number between $0$ and $100$. Denote by $m$ the mean of all strategies played. A player whose number is the closest to $pm$, where $p>0$ is the predetermined constant known to all players before the game, wins. In case of a tie, all tied players get the corresponding share of the prize. When all real numbers from $[0,100]$ are allowed, for $p\in(0,1)$ there is a unique Nash equilibrium in the model where all players choose $0$.  In a particular case of $n=2$, choosing $0$ is a weakly dominant strategy. If only integer numbers are allowed, there do exist additional Nash equilibria where the players play higher numbers. For example, if $p=0.5$, all strategy profiles where most of the players play $1$ and other players play $0$, are additional Nash equilibria. If $p=2/3$, the profile $(1,\ldots,1)$ is the only additional Nash equilibrium. 

Multiple experiments show that in the \textit{Guess the number} game people, in general, do not play Nash equilibrium. In the pioneering experimental paper, \cite{Nagel1995} demonstrated that in sessions with $p=1/2$ and $p=2/3$ no subject chose $0$ and only 6 percent chose numbers below $10$. However, in the iterated game the strategies converged to Nash equilibrium from period to period, after the participants learned statistics from the previous rounds \citep{Nagel1995}. If one uses the median of the chosen numbers instead of the mean, results do not change much in a one-shot game but in the iterated game convergence to $0$ is faster in the median variant compared to the mean variant \citep{Duffy1997}. Switching to the maximum instead of the mean or the median increases the chosen strategies \citep{Duffy1997}.

In a particular case of $n=2$, one could possibly anticipate a significant share of players choosing $0$, a weakly dominant strategy. However, this is not the case. Only $10\%$ of undergraduate students and $37\%$ of the audience of economics or psychology decision-making conferences chose $0$ \citep{Grosskopf2008}. Also, 
the mean of the numbers chosen by the professionals (22) is lower than the mean of the numbers chosen by the students (36). A higher number of participants $n=18$ leads to a lower mean both for professionals (19) and students (29). However, in the case of professionals, this difference is not statistically significant \citep{Grosskopf2008}. \cite{rydval2009three} strengthen this finding by identifying that only nearly 1/3 of all participants think in terms of dominant strategies in 2- or 3-person \textit{Guess the number}-like games with dominant strategies, while 2/3 of all participants fail to identify the strategic properties of the game. In our study, we ask whether LLMs are able to identify the strategic nature of the game.
 
Several theoretical models explaining the non-equilibrium behavior were proposed in the literature. Most of these models deal with the notion of \textit{bounded rationality} when players are rational only to some extent; the degree of rationality is associated with the sophistication of a player. A dynamical model where the players choose one of the step-$k$ behavioral rules, learn the results of the experiment, and choose more successful rules in the next iterations, was presented and estimated in \cite{Stahl1996}. A further extension of the set of possible behavioral strategies is discussed in \cite{Stahl1998}. \cite{Ho1998} builds the bounded rationality models based on the iterative deletion of dominated strategies and iterated best replies to previously played actions. It appears that many participants of experiments are using iterated best response arguments. Namely, \cite{Bosch2002} describes an experiment organized by \textit{The Financial Times} where 64\% of players indeed explained that they exploited the best responses to the revealed statistics. Note that playing iterative best response to the previous iteration of the game does not lead to the best response to other players' strategies in the new iteration \citep{Breitmoser2012}. \cite{Weber2003} demonstrated that the feedback from organizers plays a key role in the speed of convergence to Nash equilibrium: in the absence of the feedback, the numbers also decreased but at a lower rate. Advice from a peer participant has an even stronger effect on the performance than pure statistics provided by the organizers \citep{Kocher2014}. The authors of the latter paper relate it to the limited ability of players to analyze statistical data.

Cognitive ability is also an important determinant of the outcome of the \textit{Guess the number} game. Higher cognitive ability may manifest itself through better inductive reasoning, iterative dominance, and level-$k$ thinking \citep{carpenter}. \cite{Brocas2020} designed a variant of the \textit{Guess the number} game and demonstrated that the equilibrium behavior increases significantly between 5 and 10 years of age (from 17.9\% to 61.4\%) and stabilizes afterward. Back to the classic variant of the game, players with higher scores in a cognitive ability test choose lower numbers \citep{Burnham2008} and show faster convergence to equilibrium in iterated experiments \citep{Gill2016}. Mixed evidence was reported in \cite{Branas-Garza2012}: the better performance in the CRT test that measures cognitive reflectiveness is associated with lower numbers in the \textit{Guess the number} game, whereas the outcome of the Raven test measuring visual reasoning and analytic intelligence surprisingly was not associated with the successful performance in the \textit{Guess the number} game. Interestingly, high cognitive ability test scorers better respond to the cognitive ability of their opponents \citep{Gill2016}, whereas players whose abilities are below a certain threshold do not adapt strategically to the opponents' level of sophistication at all \cite{fehr}. The recent \cite{Gill2025} study distinguishes between cognitive ability and judgment. The authors show that whereas high cognitive ability shifts the strategy towards 0, high judgment subjects are less inclined to choose 0, even though their choices on average are lower than low judgment subjects. Another evidence that the level of players' sophistication matters, comes from experiments where teams consisting of several players played instead of single players. The strategies of teams of 2 players do not differ significantly from the strategies chosen by individual players \citep{Sutter2005}. At the same time, teams consisting of 3 and 4 players outperform individual players \citep{Kocher2005, Sutter2005}. One should also distinguish between cognitive abilities and cognitive effort. \cite{alos} propose an experiment where the deliberation time in the \textit{Guess the number} game serves as a proxy for cognitive effort. The authors find evidence that longer deliberation time is related to playing strategies associated with higher numbers of reasoning steps. After running experiments with two families of games, \cite{georganas} warn us that strategic sophistication varies across different families of games which prevents us from making too strong conclusions and expanding them from the \textit{Guess the number} game to more general environments.

One could hypothesize that emotions affect the players' decisions by diminishing the ability to perform deep analysis of the game. However, the evidence differs for various conditions. Players experiencing stress during the game indeed choose higher numbers \citep{Leder2013}. Angry participants of the experiment have a lower level of reasoning compared to the control group \citep{Castagnetti2023}. At the same time, sadness has little effect on the players' strategies \citep{Castagnetti2023}.

It appears that framing of the problem also matters. \cite{Hanaki2019} considered two variants of the \textit{Guess the number} game. In the first one, the players' responses are strategic complements, while in the second one their responses are strategic substitutes. Theoretical Nash equilibrium is the same in both variants. However, the authors demonstrate that the strategic environment effect manifests itself for sufficiently large or uncertain groups of players: the players' responses begin to diverge starting from 5-players games.   

There are two principal dimensions of the problem in consideration.

\begin{enumerate}
    \item For a wide range of parameters of a one-shot game, one can compare the strategies of humans and LLMs in a one-shot \textit{Guess the number} game.

    \item For a specific set of parameters, one can compare the  strategies of humans and LLMs in the iterated version of the \textit{Guess the number} game.
\end{enumerate}

While the first approach focuses on the role of changing conditions of the game, the second approach studies learning effects. Our paper follows the former one. In order to make a fair comparison, we took only the first iterations of the game from the previous literature. Similarly, we never asked LLMs to play the \textit{Guess the number} game several times in a row during one iteration of an experiment, and we never provided any feedback to LLMs. In contrast, the working paper \cite{Lu2024} follows the second approach.

For more experimental and theoretical results on \textit{Guess the number} games, we refer the reader to one of the surveys \citep{Nagel2008, Nagel2017}. \cite{fan2024} presents a detailed overview of LLMs research published by 2023. 

By performing a series of experiments, we aimed to investigate the decision-making process and strategy formulation of the modern LLMs playing against different groups of virtual players. We want to identify to what extent LLMs behave like a human player and whether LLMs can successfully identify the other players' level of sophistication. To move forward on this path, we break our task down into several questions. 

\textit{Q1. Does LLM recognize the rules of the game and act in accordance with the rules?}

This is the simplest yet essential test of LLM abilities. If an LLM fails to understand the rules and plays invalid strategies, all further investigations become meaningless.

\textit{Q2. Does LLM recognize the strategic context of the game?}

It is important to understand whether LLM takes into account the strategies of other players when choosing their own strategy. The negative answer would make us think of LLM as a very simple, unsophisticated player.

\textit{Q3. Are LLM's decisions in line with the expected comparative statics with respect to the parameters of the experiment?}

Theoretical models of bounded rationality and the empirical evidence predict that more sophisticated players behave closer to Nash equilibrium in the \textit{Guess the number} game. We aim to test how LLM responds to changes in parameters of the game. Failing to fulfill the expectations will be perceived as a disappointing signal about the LLM abilities.

\textit{Q4. Can LLM find an analytical solution to the game?}

This is an interesting question per se because neither positive nor negative answer would make the comparison of human and LLM strategies less meaningful. Recall that some of the previous experiments included those participants who are definitely unfamiliar with dominant strategies and Nash equilibrium concepts, whereas other experiments included game theorists who are presumably able to find the theoretical equilibrium in the \textit{Guess the number} game. We evaluate whether an LLM can identify a weakly dominant strategy in a two-player game.

\textit{Q5. Can LLM correctly use the analytical solution when choosing its strategy?}

In case of a positive answer, for the game of $n=2$ players we would expect that LLM always plays a weakly dominant strategy. For the $n>2$ case, we do not infer too strong conclusions because dealing with the real-world competitors requires the correct evaluation of their complexity rather than following the theoretical predictions. 

\textit{Q6. Do different LLMs perform differently?}

On the one hand, treating LLMs as black boxes means that we expect potentially different outcomes depending on what is inside those boxes. On the other hand, some tasks can be so simple (for example, asking to find the sum 2+2) that different models would provide the same output. 

\textit{Q7. Are LLM's strategies similar to strategies played by human players?}

Though this question is the most important for us, the nuances that may arise in questions \textit{Q1}--\textit{Q6} could affect the interpretation of the results. Therefore, we do not restrict our attention to \textit{Q7} solely. 

Testing LLMs with the \textit{Guess the number} game offers a dual benefit. By comparing model outputs with well-established human data, we can assess whether LLMs capture the bounded rationality and iterative reasoning typical of human decision-making. At the same time, any systematic differences highlight limitations in LLM strategic reasoning, providing valuable insights for refining these models and for their deployment in economic applications.

We experimented with the main LLMs that were available in 2024–2025: GPT-4o, GPT-4o-Mini, Gemini-2.5-flash, Claude-Sonnet-4, and Llama-4-Maverick. Apart from the comparison of human versus LLMs’ behavior, we also discuss the differences in behavior across these models.

We found that LLMs tend to see competitors as more sophisticated agents than human players do. Note that this is not a general property of LLMs. For example, in a money request game LLMs are found to be less sophisticated than human players \citep{Gao2024}.

Finally, note that we did not organize games between LLMs in this study. We refer those who are interested in tournaments between artificial players to \cite{Guo2024} who introduced EconArena environment that allows to organize various competitions between AI models. A beauty contest game is one of the first games that was implemented on EconArena.
 
The rest of the paper is organized as follows. In Section 2, we describe the methodology of our research. Section 3 presents main results and some robustness checks that are further discussed in Section 4. Section 5 concludes.

\section{The Experimental Design}


We run the experiments with the following LLMs: GPT-4o\footnote{An OpenAI model released in May 2024. OpenAI describes it as a “step towards much more natural human-computer interaction” because it accepts multimodal information and quickly analyzes it to give a response. (OpenAI official website: \url{https://openai.com/index/hello-gpt-4o/}. Retrieved December 22, 2024.)}, GPT-4o mini\footnote{A smaller version of GPT-4o, described by OpenAI as the “most cost-efficient small model”. OpenAI also claims that it outperforms other small models on such tasks as reasoning, math and coding, multimodal reasoning. (OpenAI official website: \url{https://openai.com/index/gpt-4o-mini-advancing-cost-efficient-intelligence/}. Retrieved December 22, 2024.)}, Gemini-2.5-flash\footnote{A model from Google DeepMind, described as lightweight and optimized for tasks requiring speed and efficiency. (DeepMind official website: \url{https://deepmind.google/technologies/gemini/flash/}. Retrieved December 22, 2024.)}, Claude-Sonnet-4\footnote{An Anthropic mid-sized model released May 22, 2025, offering enhanced instruction-following, coding, visual data extraction, and long-context capabilities across multimodal inputs. (Anthropic official announcement: Claude 4 release, retrieved August 27, 2025.); model family documentation highlights its usability for coding agents, reasoning, and ability to handle up to 1 million token contexts. (Anthropic model page: Claude Sonnet 4 capabilities, retrieved August 27, 2025.)}\, and Llama-4-Maverick\footnote{A model from the Llama 4 family released April 5, 2025, featuring a Mixture-of-Experts architecture (17B activated parameters, 128 experts), multimodality, multilingual support, and up to 1 million token context length. (Meta AI announcement, retrieved August 27, 2025.)}. There are several reasons for such a choice of LLMs. First, we want to know whether our results are robust to differently designed and sized LLMs: our selected models are taken from several organizations (OpenAI, Google DeepMind, Anthropic, Meta AI) and exhibit diverse architectures and design philosophies. Second, we include both proprietary models (GPT-4o, GPT-4o mini, Gemini-2.5-flash, Claude-Sonnet-4) and an open-weight, instruction-tuned model (Llama-4-Maverick), further enhancing the robustness of our findings. Third, each of the selected models is recognized for state-of-the-art performance on contemporary benchmarks, making them representative of leading capabilities in the LLM landscape.

To minimize the influence of stochastic sampling and ensure comparability across models, we fixed the temperature parameter to 1.0 for all experiments. Thus, while the default setting allows some variation in outputs, the consistency of using the same temperature across models ensures that observed differences in responses reflect the models’ reasoning processes under identical conditions, rather than arising from divergent parameter choices. The data collection process was repeated within a single session environment until 50 responses were obtained. 

The experiments were structured into 16 distinct scenarios, each characterized by a combination of factors including the type of aggregate statistic used to determine the winning number (Function), the target percentage ($p$) of the aggregate statistic, the number of players involved ($n$), and the composition of the opponent group (Opponents). These factors are identical to the settings of classic experiments with real people reported in previous literature (see Table \ref{summary} for the summary).  

\begin{sidewaystable}[htbp]
\centering
\label{experiments} 
{\begin{tabular}{ccccc p{8cm}}
\toprule
\textbf{Scenario} & \textbf{Original paper} & \textbf{n} & \textbf{p} & \textbf{Function} & \textbf{Opponents} \\

\midrule
1& \cite{Nagel1995} & 18 & 1/2 & Average & Undergraduate students of various faculties \\
2& & 18 & 2/3 & Average & Undergraduate students of various faculties \\
\midrule
3&\cite{Duffy1997} & 16 & 1/2 & Average & Undergraduate students \\
4& & 16 & 1/2 & Median & Undergraduate students \\
5& & 16 & 1/2 & Maximum & Undergraduate students \\

\midrule
6& \cite{Grosskopf2008} & 2 & 2/3 & Average &
First year undergraduate students majoring in economics \\
7& & 2 & 2/3 & Average & Audience of economics or psychology decision-making conferences\\
8& & 18 & 2/3 & Average & First year undergraduate students majoring in economics \\
9& & 18 & 2/3 & Average & Audience of economics or psychology decision-making conferences \\

\midrule
10&\cite{Branas-Garza2012} & 24 & 1/2 & Average & Individuals with high CRT score \\
11& & 24 & 1/2 & Average & Individuals with low CRT score \\
12& & 24 & 2/3 & Average & Individuals with high CRT score \\
13& & 24 & 2/3 & Average & Individuals with low CRT score \\
\midrule
14&\cite{Castagnetti2023} & 3 & 0.7 & Average & Individuals experiencing anger \\
15& & 3 & 0.7 & Average & Individuals experiencing sadness \\ 
16& & 3 & 0.7 & Average & Individuals experiencing neither anger nor sad emotions \\
 
\bottomrule
\end{tabular}}
\caption{Summary of experiments replicated in this paper with an AI player. CRT = cognitive reflection test.}
\label{summary}
\end{sidewaystable}

In each scenario, we simulate one of the human participants using the LLM. Our goal is to explore the effects of specified conditions on the model's responses and compare the AI's strategies with human strategies, given all else equal. The procedure commences by addressing the application programming interface (API) with prompts containing information regarding the experimental conditions including a description of the opponents, the values of $p$ and $n$, and the specific aggregate function. Following the delivery of prompts, the responses generated by LLM were systematically recorded, capturing not only the model’s guess regarding the number but also its underlying reasoning and conclusions. To account for variability in the responses, this procedure was iterated 50 times for each scenario. Each query was treated as a new instance to avoid any learning effects. Additionally, zero-shot prompting was employed, with no examples provided to the model, in order to capture its raw reasoning while minimizing potential biases from prior interactions or fine-tuning on related tasks. This approach was chosen to closely replicate the experimental conditions with human players described above.

The following prompt for LLMs was formulated to simulate \textit{Guess the number} game taking into account the parameters of the game described above:

\begin{quotation}
`You are playing the game. In this game, $n$ players including you simultaneously choose an integer between 0 and 100, inclusive. The winner of the game is the player whose number is closest to $p$ of the $function$ of all of the numbers chosen. You are playing against the following opponents: $Opponents$. You are in the same group as the $Opponents$, you possess the same skills, knowledge, level of intelligence, education, etc. What number will you choose, and why? Provide your answer in a JSON format with two keys: "answer\_text" (which contains the full text of the answer including the reasoning and ending with your unambiguously chosen number) and "answer\_number" (which contains only your unambiguously chosen number).'
\end{quotation}

Below, we provide an example of the GPT-4o-Mini text response to the prompt that was collected into a "answer\_text" field in scenario 6 ($n=2$, $p=2/3$, mean is the aggregate function, and the opponents are first year undergraduate students majoring in either economics, political science, law, medicine or humanities with no formal training in game theory).

\begin{quotation}
`In this game, the aim is to choose a number that is closest to 2/3 of the average of all chosen numbers. Since the opponents are first-year undergraduate students with no formal game theory training, they may choose numbers based on intuition or randomness. A common heuristic might lead them to choose around the midpoint of the range (e.g., around 50). This would mean that the average could likely be around 50, leading 2/3 of that to be approximately 33. However, if I consider that many players might overshoot or take into account that others may also think along those lines, I might choose a number lower than 33. A strategic choice could be around 22 since that would be close to a reasonable estimate of what others might average out to based on their likely choices. Therefore, I will choose 22.'
\end{quotation}

In the same experiment, the model answered with a `22' in a "answer\_number" field. 

All answers in all scenarios respected the format of data provision requested in the prompt. We accepted all answers of the models irrespectively of the potential logical drawbacks in explanations. Therefore, in our main dataset we have 50 answers from each of 5 LLMs in each of 16 scenarios.

\section{Results}

\subsection{Main Results}

This section presents the results of our analysis, comparing the performance of various AI models across 16 experimental settings derived from five previous studies of beauty contest games. We focus on the LLMs’ alignment with theoretical predictions and their consistency with observed human behavior by comparing model-generated decisions to those of human participants in the original experiments. Our findings are organized in five tables, each corresponding to a different original study.


\begin{table}[!ht]
\centering
\resizebox{\textwidth}{!}{
\begin{tabular}{cccccccccccc}
\toprule
Scenario & $n$ & $p$ & Function & Opponents & Paper & Model & Model & Model & $MM-PM$ & $t$-stat & $p$-value\\
 &  &  &  &  & mean &  & mean & st. dev. &  &  & \\
 &  &  &  &  & ($PM$) &  & ($MM$) &  &  &  & \\
\midrule
 & \multirow{5}{*}{18} & \multirow{5}{*}{1/2} & \multirow{5}{*}{Average} & \multirow{5}{*}{\shortstack{Undergraduate \\ students of various faculties}} & \multirow{5}{*}{27.05} & Claude Sonnet & 12.72 & 3.77 & -14.33 & -26.88 & 0.000 \\
 &  &  &  &  &  & GPT-4o & 20.42 & 5.76 & -6.63 & -8.14 & 0.000 \\
1 &  &  &  &  &  & GPT-4o mini & 19.58 & 5.12 & -7.47 & -10.32 & 0.000 \\
 &  &  &  &  &  & Gemini Flash & 9.17 & 3.68 & -17.88 & -34.34 & 0.000 \\
 &  &  &  &  &  & Llama & 2.00 & 3.87 & -25.05 & -45.77 & 0.000 \\
\addlinespace
 & \multirow{5}{*}{18} & \multirow{5}{*}{2/3} & \multirow{5}{*}{Average} & \multirow{5}{*}{\shortstack{Undergraduate \\ students of various faculties}} & \multirow{5}{*}{36.73} & Claude Sonnet & 16.22 & 2.57 & -20.51 & -56.52 & 0.000 \\
 &  &  &  &  &  & GPT-4o & 22.01 & 4.89 & -14.72 & -21.29 & 0.000 \\
2 &  &  &  &  &  & GPT-4o mini & 22.24 & 3.37 & -14.49 & -30.39 & 0.000 \\
 &  &  &  &  &  & Gemini Flash & 14.86 & 4.53 & -21.87 & -34.13 & 0.000 \\
 &  &  &  &  &  & Llama & 2.80 & 7.23 & -33.93 & -33.18 & 0.000 \\
\bottomrule
\end{tabular}
}
\caption{Replication results for \cite{Nagel1995} with a LLM player. In \cite{Nagel1995}, $n$ varies from 15 to 18 in different sessions. In our experiments, we fixed the number of players at 18 and assumed that the marginal effect of one additional player in the group of 15--18 players is low.}
\label{tab:nagel1995}
\end{table}

Table~\ref{tab:nagel1995} shows replicated results for the pioneering \cite{Nagel1995} experiment, which explores strategic reasoning among undergraduate students from various faculties. The table reports results for two scenarios: one with a target fraction \(p = 1/2\) and another with \(p = 2/3\), both applied to the average of all responses. In both scenarios, the AI agents demonstrate a tendency to produce guesses closer to zero, the Nash equilibrium strategy, compared to the human participants' averages reported in the paper. For instance, in the \(p = 1/2\) case, the mean guesses of all AI models are lower than the human mean of 27.05, ranging from 2.00 to 20.42. All differences with 27.05 are statistically significant at any reasonable level. For \(p = 2/3\), all AI models also tend to play closer to zero, with model means ranging from 2.80 (Llama) to 22.24 (GPT-4o mini), compared to the human mean of 36.73. Interestingly, the GPT models deviate from the other models, producing higher guesses that are closer to those of human participants, whereas Llama produces the smallest numbers. These variations are probably due to differences in how the models interpret strategic reasoning and reflect the diversity of decision-making paradigms across LLMs.

Table~\ref{tab:duffy1997} replicates findings from the \cite{Duffy1997} experiment which compares decision-making strategies using three different aggregation methods: the average, the median, and the maximum. The participants in the original study were undergraduate students. The paper reported no significant differences between the strategies in the mean and the median games while in the maximum game people choose significantly higher numbers than in either the mean or median games. For the median aggregation function, the AI agents display a range of mean guesses, with Llama providing the lowest mean guess (5.22) and GPT-4o showing the highest mean (24.26). Results under the average aggregation function in terms of models' behavior are similar to those under the median function. For the maximum aggregation function, the results reveal more variation, with Llama yielding significantly lower guesses (6.52), compared to the higher mean of 34.94 observed for Claude Sonnet. The behavior of LLM agents in the maximum aggregation condition mirrors that of human participants, with systematically higher guesses. However, while the paper's original findings reported no significant differences between the median and mean aggregation methods, the AI agents exhibit some divergence in their performance, reflecting differences in how the models process these aggregation rules. 

\begin{table}[!ht]
\centering
\resizebox{\textwidth}{!}{
\begin{tabular}{ccccccccc}
\toprule
Scenario & $n$ & $p$ & Function & Opponents & Paper result & Model  & Model mean & Model st. dev \\
\midrule

& \multirow{5}{*}{16} & \multirow{5}{*}{1/2} & \multirow{5}{*}{Average} & \multirow{5}{*}{Undergraduate students} & \multirow{5}{*}{Difference with median is not significant} 
& Claude Sonnet & 13.66 & 4.78 \\
& & & & & & GPT-4o & 19.17 & 5.08 \\
3 & & & & & & GPT-4o mini & 20.00 & 4.11 \\
& & & & & & Gemini Flash & 7.98 & 5.45 \\
& & & & & & Llama & 2.76 & 4.13 \\
\addlinespace

& \multirow{5}{*}{16} & \multirow{5}{*}{1/2} & \multirow{5}{*}{Median} & \multirow{5}{*}{Undergraduate students} & \multirow{5}{*}{Difference with mean is not significant} 
& Claude Sonnet & 12.46 & 3.75 \\
& & & & & & GPT-4o & 24.26 & 3.21 \\
4 & & & & & & GPT-4o mini & 22.94 & 3.66 \\
& & & & & & Gemini Flash & 7.06 & 3.67 \\
& & & & & & Llama & 5.22 & 6.06 \\
\addlinespace

& \multirow{5}{*}{16} & \multirow{5}{*}{1/2} & \multirow{5}{*}{Maximum} & \multirow{5}{*}{Undergraduate students} & \multirow{5}{*}{Maximum of three} 
& Claude Sonnet & 34.94 & 7.97 \\
& & & & & & GPT-4o & 30.82 & 6.58 \\
5 & & & & & & GPT-4o mini & 24.92 & 5.33 \\
& & & & & & Gemini Flash & 15.82 & 14.40 \\
& & & & & & Llama & 6.52 & 9.36 \\
\bottomrule
\end{tabular}
}
\caption{Replication results for \cite{Duffy1997} with a LLM player.}
\label{tab:duffy1997}
\end{table}

Table~\ref{tab:grosskopf2008} presents our replication of the experiment reported in \cite{Grosskopf2008}. In the original study, multiple audiences were involved, allowing for an examination of how domain knowledge and familiarity with strategic thinking might influence guesses in a beauty contest.

In Scenarios 6 and 7 (both two-player games), the original human means were 35.57 for undergraduate students and 21.73 for conference audiences (Economics/Psychology decision-making). In Scenario 6, all five models on average play significantly lower numbers than humans. In Scenario 7, GPT-4o mini plays an average of 19.54, which is not statistically different from 21.73 on 5\% level. Other models still play lower numbers than humans. A similar pattern appears in Scenarios 8 and 9, where 18-player games yielded human means of 29.31 (undergraduates) and 18.98 (game theory conference audiences). For Scenario 8, the results are consistent with those from Scenarios 6 and 7, indicating that LLM agents tend to guess below human averages. However, in Scenario 9, GPT-4o and GPT-4o mini provide statistically significantly higher guesses than human participants, with differences of 1.54 and 2.60, respectively. Overall, the results of replication of this paper align with our earlier observations that most AI models exhibit guesses closer to zero, even when human participants themselves may be relatively sophisticated (e.g., conference attendees).



\begin{table}[!ht]
\centering
\resizebox{\textwidth}{!}{
\begin{tabular}{cccccccccccc}
\toprule
Scenario & $n$ & $p$ & Function & Opponents & Paper & Model & Model & Model & $MM-PM$ & $t$-stat & $p$-value\\
& & & & & mean & & mean & st. dev. & & & \\
& & & & & ($PM$) & & ($MM$) & & & & \\
\midrule
  & \multirow{5}{*}{2} & \multirow{5}{*}{2/3} & \multirow{5}{*}{Average} & \multirow{5}{*}{\shortstack{First year undergraduate students\\ majoring in Economics}} & \multirow{5}{*}{35.57}
  & Claude Sonnet & 21.70 & 2.64 & -13.87 & -37.10 & 0.000 \\
  & & & & & & GPT-4o       & 25.96 & 5.83 &  -9.61 & -11.65 & 0.000 \\
6 & & & & & & GPT-4o mini  & 23.60 & 4.60 & -11.97 & -18.39 & 0.000 \\
  & & & & & & Gemini Flash & 22.58 & 3.39 & -12.99 & -27.07 & 0.000 \\
  & & & & & & Llama        & 18.66 & 9.99 & -16.91 & -11.97 & 0.000 \\
\addlinespace
  & \multirow{5}{*}{2} & \multirow{5}{*}{2/3} & \multirow{5}{*}{Average} & \multirow{5}{*}{\shortstack{Audience of economics or\\ psychology-decision making conferences}} & \multirow{5}{*}{21.73}
  & Claude Sonnet & 13.88 & 6.01 &  -7.85 &  -9.24 & 0.000 \\
  & & & & & & GPT-4o       & 16.52 & 9.67 &  -5.21 &  -3.81 & 0.000 \\
7 & & & & & & GPT-4o mini  & 19.54 &10.78 &  -2.19 &  -1.44 & 0.160 \\
  & & & & & & Gemini Flash &  5.54 & 5.75 & -16.19 & -19.91 & 0.000 \\
  & & & & & & Llama        &  0.88 & 4.35 & -20.85 & -33.85 & 0.000 \\
\addlinespace
  & \multirow{5}{*}{18} & \multirow{5}{*}{2/3} & \multirow{5}{*}{Average} & \multirow{5}{*}{\shortstack{First year undergraduate students\\ majoring in Economics}} & \multirow{5}{*}{29.31}
  & Claude Sonnet & 20.56 & 2.49 &  -8.75 & -24.83 & 0.000 \\
  & & & & & & GPT-4o       & 26.10 & 5.00 &  -3.21 &  -4.54 & 0.000 \\
8 & & & & & & GPT-4o mini  & 23.74 & 4.72 &  -5.57 &  -8.35 & 0.000 \\
  & & & & & & Gemini Flash & 23.32 & 2.90 &  -5.99 & -14.59 & 0.000 \\
  & & & & & & Llama        & 15.94 &10.20 & -13.37 &  -9.27 & 0.000 \\
\addlinespace
  & \multirow{5}{*}{18} & \multirow{5}{*}{2/3} & \multirow{5}{*}{Average} & \multirow{5}{*}{\shortstack{Audience of game theory conferences}} & \multirow{5}{*}{18.98}
  & Claude Sonnet & 15.34 & 5.40 &  -3.64 &  -4.77 & 0.000 \\
  & & & & & & GPT-4o       & 20.52 & 4.51 &   1.54 &   2.41 & 0.020 \\
9 & & & & & & GPT-4o mini  & 21.58 & 4.90 &   2.60 &   3.75 & 0.000 \\
  & & & & & & Gemini Flash &  7.38 & 6.10 & -11.60 & -13.44 & 0.000 \\
  & & & & & & Llama        &  0.72 & 3.58 & -18.26 & -36.06 & 0.000 \\
\bottomrule
\end{tabular}
}
\caption{Replication results for \cite{Grosskopf2008} with a LLM player.}
\label{tab:grosskopf2008}
\end{table}

Replication results for \cite{Branas-Garza2012} experiment are reported in Table~\ref{tab:branas2012}. The latter paper examines the influence of cognitive reflection on decision-making in strategic games. The original study reported that individuals with higher Cognitive Reflection Test (CRT) scores tend to choose lower numbers compared to those with lower CRT scores. Across both values of $p$, all models display lower guesses under the high-CRT scenarios (Scenarios 10 and 12) compared to the low-CRT ones (Scenarios 11 and 13), aligning well with the human data from \cite{Branas-Garza2012}.

\begin{table}[!ht]
\centering
\resizebox{\textwidth}{!}{
\begin{tabular}{c c c c c c c c c}
\toprule
Scenario & $n$ & $p$ & Function & Opponents & Paper result & Model & Model mean & Model st. dev. \\
\midrule

 & \multirow{5}{*}{24} & \multirow{5}{*}{1/2} & \multirow{5}{*}{Average} & \multirow{5}{*}{\shortstack{Individuals with high \\ cognitive reflection test (CRT) score}} & \multirow{5}{*}{\shortstack{Lower numbers compared to \\ players with low CRT scores}} 
& Claude Sonnet & 5.36 & 4.68 \\
 & & & & & & GPT-4o & 9.40 & 7.45 \\
10 & & & & & & GPT-4o mini & 16.02 & 6.30 \\
 & & & & & & Gemini Flash & 0.12 & 0.39 \\
 & & & & & & Llama & 0.22 & 1.02 \\
\addlinespace

 & \multirow{5}{*}{24} & \multirow{5}{*}{1/2} & \multirow{5}{*}{Average} & \multirow{5}{*}{\shortstack{Individuals with low \\ cognitive reflection test (CRT) score}} & \multirow{5}{*}{\shortstack{Higher numbers compared to \\ players with high CRT scores}} 
& Claude Sonnet & 21.00 & 4.87 \\
 & & & & & & GPT-4o & 22.76 & 3.73 \\
11 & & & & & & GPT-4o mini & 19.36 & 4.67 \\
 & & & & & & Gemini Flash & 16.58 & 4.50 \\
 & & & & & & Llama & 12.60 & 7.01 \\
\addlinespace

 & \multirow{5}{*}{24} & \multirow{5}{*}{2/3} & \multirow{5}{*}{Average} & \multirow{5}{*}{\shortstack{Individuals with high \\ cognitive reflection test (CRT) score}} & \multirow{5}{*}{\shortstack{Lower numbers compared to \\ players with low CRT scores}} 
& Claude Sonnet & 8.48 & 5.53 \\
 & & & & & & GPT-4o & 8.60 & 8.82 \\
12 & & & & & & GPT-4o mini & 19.38 & 4.76 \\
 & & & & & & Gemini Flash & 0.38 & 1.26 \\
 & & & & & & Llama & 0.46 & 3.11 \\
\addlinespace

 & \multirow{5}{*}{24} & \multirow{5}{*}{2/3} & \multirow{5}{*}{Average} & \multirow{5}{*}{\shortstack{Individuals with low \\ cognitive reflection test (CRT) score}} & \multirow{5}{*}{\shortstack{Higher numbers compared to \\ players with high CRT scores}} 
& Claude Sonnet & 27.88 & 4.58 \\
 & & & & & & GPT-4o & 30.44 & 4.75 \\
13 & & & & & & GPT-4o mini & 23.86 & 5.57 \\
 & & & & & & Gemini Flash & 24.92 & 5.46 \\
 & & & & & & Llama & 23.10 & 6.12 \\
\bottomrule
\end{tabular}
}
\caption{Replication results for \cite{Branas-Garza2012} with a LLM player.}
\label{tab:branas2012}
\end{table}

Table~\ref{tab:castagnetti2023} replicates the experiment from \cite{Castagnetti2023} who investigate the impact of emotions, particularly anger and sadness, on decision-making in strategic games. The original paper concluded that individuals experiencing anger performed less optimally compared to a control group. At the same time, players who experience sad emotions, do not play significantly different strategies compared to the conrol group. We evaluate the performance of AI agents under similar conditions.

Three scenarios are reported. In Scenario 14, participants experienced anger, and the five AI models consistently generate higher guesses (ranging from about 5.28 to 35.64) than they do in Scenario 16, the neutral control condition (where model means range from 0.02 to 29.78). These findings parallel the original conclusion that anger increases guesses.

In contrast, Scenario 15 induces sadness; here, four of the models’ guesses were closer to those in the neutral condition, suggesting that sadness at least has a weaker effect on optimality than anger. The only exception was Gemini Flash, which increased its guesses when playing against sad opponents compared to angry opponents.

\begin{table}[!ht]
\centering
\resizebox{\textwidth}{!}{
\begin{tabular}{ccccccccc}
\toprule
Scenario & $n$ & $p$ & Function & Opponents & Paper result & Model & Model mean & Model st. dev. \\
\midrule

 & \multirow{5}{*}{3} & \multirow{5}{*}{0.7} & \multirow{5}{*}{Average} & \multirow{5}{*}{\shortstack{Individuals experiencing \\ anger}} & \multirow{5}{*}{\shortstack{Less optimal compared to \\ control group}} 
& Claude Sonnet & 25.38 & 6.25 \\
 & & & & & & GPT-4o & 31.92 & 8.66 \\
14 & & & & & & GPT-4o mini & 35.64 & 9.55 \\
 & & & & & & Gemini Flash & 5.28 & 8.08 \\
 & & & & & & Llama & 10.06 & 15.59 \\
\addlinespace

 & \multirow{5}{*}{3} & \multirow{5}{*}{0.7} & \multirow{5}{*}{Average} & \multirow{5}{*}{\shortstack{Individuals experiencing \\ sadness}} & \multirow{5}{*}{\shortstack{No evidence of less optimal play \\ compared to control group}} 
& Claude Sonnet & 15.28 & 4.54 \\
 & & & & & & GPT-4o & 30.32 & 6.78 \\
15 & & & & & & GPT-4o mini & 22.52 & 8.99 \\
 & & & & & & Gemini Flash & 13.74 & 8.33 \\
 & & & & & & Llama & 2.40 & 6.60 \\
\addlinespace

 & \multirow{5}{*}{3} & \multirow{5}{*}{0.7} & \multirow{5}{*}{Average} & \multirow{5}{*}{\shortstack{Individuals experiencing neither \\ anger nor sadness}} & \multirow{5}{*}{--} 
& Claude Sonnet & 9.80 & 6.40 \\
 & & & & & & GPT-4o & 23.36 & 10.42 \\
16 & & & & & & GPT-4o mini & 29.78 & 7.98 \\
 & & & & & & Gemini Flash & 0.04 & 0.20 \\
 & & & & & & Llama & 0.02 & 0.14 \\
\bottomrule
\end{tabular}
}
\caption{Replication results for \cite{Castagnetti2023} with a LLM player.}
\label{tab:castagnetti2023}
\end{table}

\subsection{Llama Results}

In addition, a supplementary analysis was conducted on Llama models of varying sizes with respect to their number of parameters. The results for the Llama family of models clearly demonstrate that the number of parameters strongly influences performance in the beauty contest game (Table ~\ref{tab:llama_results}). As shown in the table, smaller Llama models (e.g., 1B, 3B, 8B) tend to produce guesses that deviate substantially from the theoretical Nash equilibrium of zero, often generating values comparable to or even higher than human participants’ choices. By contrast, larger versions of Llama (70B and especially 405B) yield results much closer to the equilibrium. This systematic pattern suggests that model size may serve as a reliable proxy for sophistication: the more parameters the model possesses, the closer its strategic reasoning approaches the theoretical benchmark. In other words, scaling up the Llama architecture appears to enhance its capacity for iterative reasoning, thereby making the model “smarter” in the sense of converging more closely toward equilibrium behavior.

\begin{table}[ht]
\centering
\makebox[\textwidth][c]{%
\begin{tabular}{ccccccc}
\toprule
Paper & Scenario & 1B & 3B & 8B & 70B & 405B \\
\midrule
Nagel (1995)               & 1  & 49.29 & 41.01 & 46.86 & 28.06 & 20.80 \\
Nagel (1995)               & 2  & 53.00 & 38.72 & 42.50 & 34.08 & 17.66 \\
Duffy \& Nagel (1997)      & 3  & 52.87 & 42.64 & 61.12 & 27.52 & 22.10 \\
Duffy \& Nagel (1997)      & 4  & 44.01 & 37.56 & 36.44 & 25.88 & 22.48 \\
Duffy \& Nagel (1997)      & 5  & 46.38 & 64.39 & 74.68 & 42.26 & 34.14 \\
Grosskopf \& Nagel (2008)  & 6  & 44.16 & 39.89 & 49.72 & 40.20 & 30.68 \\
Grosskopf \& Nagel (2008)  & 7  & 50.57 & 42.59 & 46.47 & 39.10 &  8.42 \\
Grosskopf \& Nagel (2008)  & 8  & 49.78 & 43.26 & 43.37 & 35.52 & 29.36 \\
Grosskopf \& Nagel (2008)  & 9  & 63.64 & 44.50 & 55.86 & 33.10 & 11.59 \\
Branas-Garza et al. (2012) & 10 & 40.12 & 50.09 & 44.66 & 23.31 & 12.91 \\
Branas-Garza et al. (2012) & 11 & 38.60 & 49.68 & 45.40 & 28.43 & 27.12 \\
Branas-Garza et al. (2012) & 12 & 49.50 & 39.40 & 44.03 & 30.30 &  9.66 \\
Branas-Garza et al. (2012) & 13 & 47.41 & 35.79 & 43.28 & 34.96 & 32.10 \\
Castagnetti et al. (2023)  & 14 & 25.92 & 32.62 & 35.37 & 40.60 & 39.62 \\
Castagnetti et al. (2023)  & 15 & 26.18 & 29.64 & 34.00 & 40.48 & 27.54 \\
Castagnetti et al. (2023)  & 16 & 26.64 & 34.55 & 38.26 & 35.58 & 22.90 \\
\bottomrule
\end{tabular}}
\caption{Average predictions of Llama models of different sizes, from 1 billion parameters to 405 billion parameters.}
\label{tab:llama_results}
\end{table}

\subsection{Robustness checks}

To ensure the stability of our findings, three robustness checks were performed by systematically modifying the experimental prompts. 

First, the original prompt was paraphrased using synonyms, resulting in the following alternative formulation:

\begin{quotation}
`Picture yourself in a contest with $n$ participants (you included). Everyone simultaneously picks an integer from 0 to 100, inclusive. Once all numbers are in, a $function$ is applied to the full set; its value at $p$ is computed. The victor is the player whose chosen number lies nearest to that computed value.
Your rivals are: $Opponents$. Assume you and they share identical ability, knowledge, intelligence, and education.
Task: Which integer will you choose, and why?
Return your reply in JSON with exactly two keys: 'answer\_text' (which contains the full text of the answer including  the reasoning and ending with your unambiguously chosen number) and 'answer\_number' (which contains only your unambiguously chosen number.'
\end{quotation}

\begin{table}[ht]
\small
\centering
\makebox[\textwidth][c]{%
\begin{tabular}{ccccccc}
\toprule
Paper & Scenario & Claude Sonnet & Gemini Flash & GPT-4o & GPT-4o mini & Llama \\
\midrule
Nagel (1995)               & 1  & 11.72 &  5.68 & 20.30 & 24.80 &  9.84 \\
Nagel (1995)               & 2  & 16.96 &  7.46 & 18.72 & 33.52 &  6.60 \\
Duffy \& Nagel (1997)      & 3  & 10.62 &  3.76 & 20.64 & 25.08 & 17.78 \\
Duffy \& Nagel (1997)      & 4  &  8.42 &  1.16 & 24.98 & 26.44 & 17.76 \\
Duffy \& Nagel (1997)      & 5  & 27.50 & 12.90 & 40.08 & 48.98 & 34.46 \\
Grosskopf \& Nagel (2008)  & 6  & 18.88 & 19.74 & 23.32 & 32.04 &  8.38 \\
Grosskopf \& Nagel (2008)  & 7  & 13.74 &  1.04 & 16.30 & 30.26 &  2.24 \\
Grosskopf \& Nagel (2008)  & 8  & 16.80 & 22.56 & 22.58 & 32.40 & 13.24 \\
Grosskopf \& Nagel (2008)  & 9  & 13.16 &  2.48 & 20.62 & 28.54 &  3.98 \\
Branas-Garza et al. (2012) & 10 &  2.70 &  0.00 & 10.88 & 25.18 &  3.50 \\
Branas-Garza et al. (2012) & 11 & 14.92 & 19.38 & 23.06 & 25.72 & 21.28 \\
Branas-Garza et al. (2012) & 12 &  2.80 &  0.10 & 15.38 & 31.76 &  6.88 \\
Branas-Garza et al. (2012) & 13 & 21.30 & 22.32 & 28.32 & 30.74 & 23.92 \\
Castagnetti et al. (2023)  & 14 & 19.86 & 15.80 & 31.36 & 40.40 & 21.24 \\
Castagnetti et al. (2023)  & 15 & 13.42 &  1.50 & 30.18 & 32.28 & 15.72 \\
Castagnetti et al. (2023)  & 16 & 15.10 &  0.76 & 23.26 & 35.16 & 13.06 \\
\bottomrule
\end{tabular}}
\caption{Robustness checks for paraphrased prompt using synonyms.}
\label{tab:robust_synonyms}
\end{table}

A comparison of the results generated under the original and synonym-based prompts (Table~\ref{tab:robust_synonyms}) indicates that the overall behavioral patterns are preserved. 

The results from \cite{Nagel1995} and \cite{Grosskopf2008} were compared in terms of AI agents’ behavior relative to human outcomes under varying prompts. In the experimental setting of \cite{Nagel1995}, AI agents produced similar outcomes across both prompts. In the setting of \cite{Grosskopf2008}, however, an exception emerged for GPT-4o mini, which produced different outcomes in Scenarios 7 and 8, with predicted means exceeding those of human participants. In the case of \cite{Duffy1997}, the outcomes showed that AI agents consistently produced the highest values under both prompts in the setting of maximum function.

The divergent behavior of GPT-4o mini was also observed in the context of 
\cite{Branas-Garza2012}. In this case, results of AI agents were compared between the original and synonym-based prompts across conditions defined by participants’ CRT scores and the value of the $p$ parameter. The only discrepancy occurred in Scenarios 12 and 13, where GPT-4o mini generated a slightly higher mean predictions for participants with high CRT scores compared to those with low CRT scores.

Finally, for \cite{Castagnetti2023}, model predictions were compared across conditions with emotions (anger and sadness) versus the control group. For anger, the results were consistent: AI agents exposed to anger produced higher average values under both prompts. For sadness, four out of five models (all except GPT-4o mini) produced higher values than the control group under the original prompt. Under the synonym-based prompt, however, an additional model (Claude Sonnet) also produced lower values under sadness compared to the control.

Second, to avoid direct associations with the well-known beauty contest framing, the task was reformulated in the context of a game show scenario:

\begin{quotation}
`You are a participant in a TV show, and it’s time for the next challenge. There are $n$ contestants, including you. Only one will advance to the next stage of the show. The organizers have given each participant a set of 100 stones. Simultaneously, each participant chooses how many stones (between 0 and 100) to put into the pot. The participant whose number of stones is closest to $p$ times a $function$ of all the submitted numbers of stones is declared the winner and advances to the next round. Your opponents are: $Opponents$ You are in the same group as the $Opponents$, you possess the same skills, knowledge, level of intelligence, education, etc. How many stones will you put, and why? Provide your answer in a JSON format with two keys: “answer\_text” (which contains the full text of the answer including the reasoning and concluded by your unambiguously chosen number of stones) and “answer\_number” (which contains only your unambiguously chosen number of stones).'
\end{quotation}

\begin{table}[ht]
\centering
\makebox[\textwidth][c]{%
\begin{tabular}{ccccccc}
\toprule
Paper & Scenario & Claude Sonnet & Gemini Flash & GPT-4o & GPT-4o mini & Llama \\
\midrule
Nagel (1995)               & 1  & 10.48 &  6.10 & 21.68 & 23.24 & 0.00 \\
Nagel (1995)               & 2  & 17.00 & 13.60 & 22.62 & 28.98 & 0.00 \\
Duffy \& Nagel (1997)      & 3  & 11.76 &  3.44 & 20.30 & 23.42 & 0.00 \\
Duffy \& Nagel (1997)      & 4  & 13.04 &  5.68 & 24.92 & 25.32 & 0.00 \\
Duffy \& Nagel (1997)      & 5  & 27.92 & 41.22 & 30.90 & 26.94 & 0.96 \\
Grosskopf \& Nagel (2008)  & 6  & 22.30 & 19.50 & 28.90 & 29.72 & 0.88 \\
Grosskopf \& Nagel (2008)  & 7  & 15.54 &  1.22 & 19.88 & 27.56 & 0.00 \\
Grosskopf \& Nagel (2008)  & 8  & 21.82 & 21.08 & 26.44 & 29.48 & 4.02 \\
Grosskopf \& Nagel (2008)  & 9  &  5.96 &  0.02 &  7.64 & 25.50 & 0.00 \\
Branas-Garza et al. (2012) & 10 &  3.94 &  0.00 & 13.64 & 23.18 & 0.44 \\
Branas-Garza et al. (2012) & 11 & 17.94 & 16.48 & 24.12 & 23.54 & 8.10 \\
Branas-Garza et al. (2012) & 12 &  5.12 &  0.20 &  7.78 & 26.62 & 0.00 \\
Branas-Garza et al. (2012) & 13 & 28.42 & 26.68 & 30.02 & 32.00 & 19.46 \\
Castagnetti et al. (2023)  & 14 & 39.40 & 18.72 & 36.44 & 36.66 & 0.00 \\
Castagnetti et al. (2023)  & 15 & 18.88 &  7.52 & 32.28 & 28.82 & 0.00 \\
Castagnetti et al. (2023)  & 16 & 15.08 &  1.42 & 34.84 & 34.50 & 0.00 \\
\bottomrule
\end{tabular}}
\caption{Robustness checks for paraphrasing using a different context.}
\label{tab:robust_paraphrasing}
\end{table}

The results obtained under the TV show framing broadly confirm the robustness of the initial findings, with models preserving their relative positioning in terms of strategic sophistication, although some quantitative differences arise (Table~\ref{tab:robust_paraphrasing}). For the settings examined by 
\cite{Nagel1995}, \cite{Duffy1997}, and \cite{Branas-Garza2012}, no differences were observed in model behavior between the original and paraphrased prompts.

In the setting of \cite{Grosskopf2008}, GPT-4o mini again demonstrated sensitivity to prompt paraphrasing. Under the original prompt, this model produced higher predicted values than human participants only in Scenario 9. Under the TV show prompt, however, it also produced higher values in Scenarios 7 and 8. Another instance of deviation under the paraphrased prompt was observed in Scenarios 14 and 15 of \cite{Castagnetti2023}. A notable change in behavior occurred with the Llama model, which produced zero predictions in 10 out of 16 scenarios, although it had also demonstrated lower predictions compared to other models under the original prompt. All other results from the robustness check with the paraphrased prompt remained consistent with those obtained under the original formulation.

Third, to examine the sensitivity of responses to additional contextual features, the original prompt was extended by explicitly assigning the LLM player psychological and cognitive characteristics. Specifically, the agent was described as experiencing either anger or sadness, or alternatively, as possessing either a low or high cognitive reflection test (CRT) score. We therefore added an explicit clause to the prompts, for both CRT scores and emotions during robustness checks. For example:

\begin{quotation}

\textit{You are playing the game. In this game, 3 players, including you, simultaneously choose an integer between 0 and 100, inclusive. The winner is the player whose number is closest to 0.7 times the average of all numbers chosen. You are playing against the following opponents: individuals experiencing anger. You are in the same group as the individuals experiencing anger; you possess the same skills, knowledge, intelligence, and education. \textbf{You are also experiencing anger.} Provide your answer in a JSON format with two keys: “answer\_text” (which contains the full text of the answer including the reasoning and concluded by your unambiguously chosen number of stones) and “answer\_number” (which contains only your unambiguously chosen number of stones).}
\end{quotation}

Thus, we explicitly assigned the relevant trait (CRT score or emotional state) to the model’s persona.

When emotional states and cognitive reflection traits were incorporated clearer into the prompts, model predictions remained broadly stable (Table~\ref{tab:robust_crt_emotions}). The inclusion of CRT level or emotional state had little influence on the mean values, as the variance of mean predictions was nearly identical between the extended prompt (0–37.42) and the original prompt (0.02–35.64).

With respect to CRT score \citep{Branas-Garza2012}, the finding that participants with higher CRT levels tend to play closer to 0 holds for both prompts, across conditions where the $p$ parameter equals 1/2 and 2/3. Regarding emotional states \citep{Castagnetti2023}, the behavior of AI agents with explicitly assigned emotions also indicates that participants experiencing anger or sadness provide higher values compared to the control group. On average, those in the anger condition produce higher guesses than those in the sadness condition for both prompts.

\begin{table}[ht]
\centering
\small
\makebox[\textwidth][c]{%
\begin{tabular}{ccccccc}
\toprule
Paper & Scenario & Claude Sonnet & Gemini Flash & GPT-4o & GPT-4o mini & Llama \\
\midrule
Branas-Garza et al. (2012) & 10 &  5.54 &  0.04 &  9.10 & 17.16 &  0.06 \\
Branas-Garza et al. (2012) & 11 & 23.30 & 23.82 & 23.94 & 21.36 & 24.84 \\
Branas-Garza et al. (2012) & 12 &  7.40 &  0.12 &  6.38 & 20.68 &  0.04 \\
Branas-Garza et al. (2012) & 13 & 32.50 & 29.50 & 31.98 & 26.22 & 33.40 \\
Castagnetti et al. (2023)  & 14 & 25.46 &  8.68 & 31.46 & 37.42 & 23.52 \\
Castagnetti et al. (2023)  & 15 & 20.66 & 13.88 & 31.82 & 26.84 &  3.24 \\
Castagnetti et al. (2023)  & 16 &  8.78 &  0.24 & 27.49 & 36.88 &  0.00 \\
\bottomrule
\end{tabular}
}
\caption{Robustness checks for CRT level and emotions}
\label{tab:robust_crt_emotions}
\end{table}

Taken together, the three robustness checks demonstrate that although numerical outcomes may vary under paraphrasing, alternative framings, or the introduction of psychological attributes, the relative ordering and qualitative interpretation of model performance remain consistent, underscoring the reliability of the main results. The highest sensitivity to prompt formulation is observed for GPT-4o mini and Llama. It should also be noted that human participants likewise might be sensitive to the framing of the game, and the models exhibit a comparable responsiveness.

\section{Discussion}

In the Introduction, we formulated 7 questions aimed at understanding the strategic reasoning of artificial intelligence. Now we are ready to answer those questions based on the results of our \textit{Guess the number} experiments. For each question, we provide a formal criterion for deriving an answer.

\textit{Q1. Does LLM recognize the rules of the game and act in accordance with the rules?} \textbf{(Yes.)}

Criterion: the share of correct answers, i.e. integer numbers from 0 to 100. The higher the share of correct answers, the better LLM recognizes the rules of the game and act in accordance with the rules.

In our dataset, 100\% of LLM answers are legitimate. Therefore, we conclude that LLMs do recognize and adhere to the rules.

\textit{Q2. Does LLM recognize the strategic context of the game?} \textbf{(Yes.)}

We asked GPT-4o to classify all answers into 6 categories using the following prompt:

\begin{quote}
    
Task: You are given a player's answer from the ``Keynesian Beauty Contest" game. Each player picks a number between 0 and 100 and explains their reasoning. Classify the reasoning into one of six categories:

1. No explanation.

2. Uninformative or purely heuristic (e.g., ``Because I said so", ``It’s sunny in Paris, so 47", ``3 is a lucky number, so 33").

3. Theoretical Nash equilibrium (explicitly states everyone should choose 0, so they choose 0).

4. Level-1 reasoning (expects average/median $\approx$ 50, picks fraction once).

5. Finite level-k reasoning (iterates reasoning steps $>$1 but finite times).

6. Infinite level-k reasoning (iterates reasoning to infinity, leading to 0).

Output format:
$<$category number$>$ — $<$brief explanation why$>$

The player's answer to classify is:
\end{quote}

Criterion: the share of answers falling into categories 3 -- 6 is now regarded as a degree of recognition of the strategic context of the game.

Table \ref{tab:stratclassif} presents the results of the strategy classification.

\begin{table}[ht]
\centering
\small
\makebox[\textwidth][c]{%
\begin{tabular}{cccccc}
\toprule
Strategy category & Claude Sonnet & Gemini Flash & GPT-4o & GPT-4o mini & Llama \\
\midrule
Category 1 & 0 & 0 & 0 & 0 & 0 \\
Category 2 & 0 & 0 & 4 & 19 & 0 \\
Category 3 & 0 & 20 & 11 & 1 & 91 \\
Category 4 & 96 & 45 & 339 & 396 & 45 \\
Category 5 & 688 & 527 & 403 & 373 & 268 \\
Category 6 & 16 & 208 & 43 & 11 & 396 \\
\bottomrule
\end{tabular}
}
\caption{Classification of strategy explanations.}
\label{tab:stratclassif}
\end{table}

As one can see, almost all strategies played by LLMs (except 23) included strategic level-$k$ reasoning patterns or theoretical equilibrium play. Therefore, we confirm the positive answer to question Q2: LLMs do recognize the strategic context of the game.

\textit{Q3. Are LLM's decisions in line with the expected comparative statics with respect to the parameters of the experiment?} \textbf{(Yes.)}

For each LLM, we calculate whether

\begin{enumerate}
    \item Their average response in Scenario 1 is lower than in Scenario 2;

    \item Their average response in Scenario 7 is lower than in Scenario 6;

    \item Their average response in Scenario 9 is lower than in Scenario 8;

    \item Their average response in Scenario 10 is lower than in Scenario 11;

    \item Their average response in Scenario 12 is lower than in Scenario 13;

    \item Their average response in Scenario 16 is lower than in Scenario 14.
    
\end{enumerate}

Criterion: the higher is the number of positive answers, the better LLM's decisions are in line with the expected comparative statics with respect to the parameters of the experiment.

Based on Tables \ref{tab:nagel1995} -- \ref{tab:castagnetti2023}, we report that each of five LLMs has 6 positive answers out of 6. Therefore, all models demonstrate expected adaptability to the changing set of parameters in our experiments.    

\textit{Q4. Can LLM find an analytical solution to the game?} \textbf{(No.)}

For scenarios with $n=2$, we asked GPT-4o to classify all answers into 2 categories using the following prompt:

\begin{quotation}
    Task: You are given a player’s answer from the Keynesian Beauty Contest game with 2 players. Each player chooses a number between 0 and 100 and explains their reasoning. In this game, playing 0 is a weakly dominant strategy.

Your task is to classify the player’s reasoning into one of two categories:

1. Dominant strategy explanation. The reasoning explicitly explains that playing 0 is a weakly dominant strategy. It is not enough to simply mention that 0 is a Nash equilibrium. To belong to this category, the reasoning must include some explanation that playing 0 is weakly better regardless of what the opponent chooses (Example answer in this category: “No matter what the other player picks, playing 0 is never worse and sometimes strictly better, so 0 is the weakly dominant strategy. I will play 0.”).

2. Other explanations. All other reasoning that does not meet the criteria for category 1. (Example answer in this category: “If both players are rational, they will both choose 0 since it is a Nash equilibrium.”).

The player’s answer to classify is:
\end{quotation}

Criterion: the share of answers falling into category 1 shows how well LLMs identify weakly dominant strategies.

We report that none of the responses were classified as category 1. LLMs fail to identify the dominant strategy in the \textit{Guess the number} game with two players.

\textit{Q5. Can LLM correctly use the analytical solution when choosing its strategy?} \textbf{(Indefinite.)}

Deriving the dominant strategy does not guarantee that an LLM will actually play this strategy.

Criterion: for scenarios with $n=2$, we compute the share of cases when LLM played $0$ in responses that were classified as belonging to category 1 in Q5. This criterion is not applicable if no responses were classified as category 1 in Q5.

Since no strategies were classified as category 1 in Q5, we cannot assess whether LLMs are able to correctly apply the concept of a dominant strategy.

\textit{Q6. Do different LLMs perform differently?} \textbf{(Yes.)}

For each particular scenario and any given pair of LLMs we conduct a $t$-test for comparing the average numbers of the corresponding LLMs.  

Criterion: if the test does not reject a hypothesis of equal averages at the 1\% significance level, we give a negative response to Q6 (for the particular scenario and particular pair of LLMs).

For the sake of brevity, we do not provide all calculations that can be easily reproduced based on Tables \ref{tab:nagel1995}--\ref{tab:castagnetti2023}. While some model pairs (18\% of all pairs in all 16 scenarios) exhibit statistical equivalence, the results show systematic deviations, suggesting that the models adopt distinct strategies in generating numerical choices. This increased heterogeneity underscores the importance of model architecture and training background in shaping behavioral outputs.

\textit{Q7. Are LLM's strategies similar to strategies played by human players?} \textbf{(No.)}.

For each of scenarios 1, 2, 6, 7, 8, and 9 (i.e., the scenarios in which the exact numbers played by humans are known), we conduct a $t$-test to compare the average numbers of humans and LLMs for any given model. 

Criterion: if the test does not reject a hypothesis of equal averages at the 1\% significance level, we give a negative response to Q7 (for the particular scenario and particular LLM).

The comparison indicates that, in most experimental settings, model-generated averages significantly diverge from human results, with exact matches being relatively rare (7\% of all comparisons in 6 scenarios).

\section{Conclusion}

This study explored the performance of various LLMs in replicating human-like strategic reasoning across a range of experimental settings derived from five behavioral studies dealing with the \textit{Guess the number} game. By replicating the experiments with such models as Claude Sonnet, Gemini Flash, GPT-4o, GPT-4o Mini, and Llama, we assessed their ability to align with human decision-making patterns, adapt to varying experimental parameters, and approximate theoretical predictions.

Our results show that LLMs recognize strategic context of the game and demonstrate expected adaptability to the changing set of parameters. LLMs systematically behave in a more sophisticated way compared to the participants of the original experiments (they play lower numbers compared to the numbers played by the human players in a similar setting). All models failed to play a dominant strategy in a two-player game. These results suggest that while some LLMs are capable of emulating nuanced strategic behavior, their responses are often shaped by their underlying architectures and design priorities. 

A limitation of our study is that we did not incorporate an explicit payoff structure in the AI prompts. The original experiments featured varying incentive mechanisms, which may have influenced participants’ choices. Our decision to leave incentives unspecified—while aiming to replicate original conditions—might have affected the observed strategic behavior of the LLMs. Future research should investigate how different incentive specifications influence AI decision-making in strategic settings.

Another limitation is that the alternative to use the same instructions that humans got in the respective experiments, is not feasible. The context matters. For example, participants of a game theory conference understand who they are. In contrast, LLMs need additional information about their competitors and themselves in the prompt. Therefore, we stick to the second-best option of using prompts that are as similar to the original instructions as possible.

Finally, attributing fixed traits to LLMs cannot fully capture the endogenous nature of human characteristics (e.g., CRT rates, sad emotions). Nevertheless, researches use persona/role prompting to study trait-dependent behavior in LLMs. For example, \cite{kroczek2025influence} assigned extroverted versus introverted roles to an LLM agent and showed that both its outputs and people’s willingness to seek help vary with the assigned role. Other examples of this methodology may be found in \cite{de2024use,jiang2023personallm,wang2025evaluating}. Taken together, these studies indicate that while LLMs are not humans, persona conditioning is a widely used and reproducible method for probing hypothesized mechanisms, and thus provides a justified complement to human-participant research.

This study contributes to the growing body of research on the potential of LLMs to model human behavior in economic decision-making contexts. Future work could extend these analyses to more complex strategic environments, incorporate additional behavioral datasets, and explore ways to enhance the adaptability of LLMs to further bridge the gap between artificial and human decision-making. The convergence of AI's calculated rationality with human intuition and behavior would open new avenues for enhancing predictive models and designing economic policies that account for the bounded rationality in human economic activities.

\section*{Acknowledgements}

We thank the editor and the three anonymous reviewers for their helpful comments, which substantially improved the paper. We are also grateful to the participants of the European Meeting on Game Theory 2024 (SING19) and the HSE International Laboratory of Game Theory and Decision Making research seminar for their valuable feedback and suggestions.

\section*{Funding}

The authors gratefully acknowledge support from the Basic Research Program of the HSE University. The funding source had no involvement in the research.

\section*{Declaration of generative AI and AI-assisted technologies in the writing process}

During the preparation of this work the authors used GPT-4o, GPT-4o mini, Gemini-1.5-flash, Claude-3.5-Sonnet, and Llama-3.1-8B-Instruct-Turbo as economic agents in order to run the experiments. Also, GPT-4o was used for language proofreading purposes. After using these tools, the authors reviewed and edited the content as needed and take full responsibility for the content of the published article.

\section*{Data availability}

The data supporting the findings of this study is publicly available in Harvard Dataverse: \url{https://doi.org/10.7910/DVN/LURKWJ}.

\bibliographystyle{apa-good}
\bibliography{sample-bibliography}

\end{document}